\documentstyle[aps,prl,twocolumn,rotating,epsfig]{revtex}
\begin{document}
\draft
\preprint{}

\title{Spin effects in the magneto-drag between double quantum wells}

\author{J.G.S. Lok$^1$, S. Kraus$^1$, M. Pohlt$^1$, W. Dietsche$^1$, K. von Klitzing$^1$, W. Wegscheider$^{2,3}$, and M. Bichler$^2$}

\address{$^1$Max-Planck-Institut f\"ur Festk\"orperforschung, Heisenbergstrasse 1, 70569 Stuttgart, Germany\\
$^2$Walter Schottky Institut, Technische Universit\"at M\"unchen, 85748 Garching, Germany\\
$^3$Institut f\"ur angewandte und experimentelle Physik, Universit\"at Regensburg, 93040 Regensburg, Germany}

\date{2 April 2000}

\maketitle
\begin{minipage}[h]{17.5cm}
\vspace*{-0.8cm}
\begin{abstract}
\noindent
We report on the selectivity to spin in a drag measurement. This selectivity
to spin causes deep minima in the magneto-drag at odd fillingfactors
for matched electron densities at magnetic fields and temperatures at which
the bare spin
energy is only one tenth of the temperature. For mismatched densities
the selectivity causes a novel 1/B-periodic oscillation, such that
{\em negative} minima in the drag are observed whenever the majority spins at
the Fermi energies of the two-dimensional electron gasses (2DEGs) are
{\em anti-parallel}, and {\em positive} maxima whenever the majority spins at
the Fermi energies are {\em parallel}.
\end{abstract}
\vspace*{-0.2cm}
\pacs{PACS: 72.80.E, 73.20.Dx, 73.40.Hm}
\end{minipage}
\vspace*{-0.1cm}
\narrowtext

\par
The physics of two-dimensional electron gases (2DEGs) has spawned
numerous discoveries over the last two decades with the integer and
fractional quantum Hall effects being the most prominent examples. More
recently, interaction phenomena between closely spaced 2DEGs in
quantizing magnetic fields have found strong interest both
experimentally~\cite{Pellegrini,Sawada,Khrapai} and
theoretically~\cite{Zheng,Brey}, because of the peculiar role the
electron spin plays in these systems. Particularly interesting is a
measurement of the frictional drag between two 2DEGs, as it probes the
density-response functions in the limit of low frequency and finite
wavevector (see~\cite{Rojo} and references therein), \hbox{a quantity which
is not easily accessible otherwise.}
\par
Experimental data of drag at zero magnetic field are
reasonably well understood. Several puzzling issues
however exist for the magneto-drag. Firstly, at matched densities in
the 2DEGs, the magneto-drag displays a double peak around odd
fillingfactor~\cite{Rubel2,Hill}
when spin-splitting is not visible at all in the longitudinal resistances
of each individual 2DEG. These double peaks were ascribed to either an
enhanced screening when the Fermi energy ($E_F$) is in the centre of a
Landau level~\cite{Bonsager,Rubel2}, or to an enhanced
spin-splitting~\cite{Hill}. Secondly, at mismatched densities negative
magneto-drag has been observed~\cite{Gramila2}, i.e. an accelleration
of the electrons opposite to the direction of the net transferred momentum.
This negative drag was speculatively ascribed to a hole-like-dispersion
in the less-than-half-filled Landau levels brought about by
disorder~\cite{Gramila2}.
\par
In this Letter we present data taken in a hitherto unexplored
temperature-magnetic field regime which clearly demonstrate the decisive role
the electron spin plays in the drag. We find that {\em both} the above issues
have a common origin; they are caused by the
fact that the drag is selective to the spin of the electrons, such
that electrons with anti-parallel spin in each 2DEG have a negative
and those with parallel spin have a positive contribution
to \vbox{\noindent the drag. At mismatched densities this selectivity causes
a novel 1/B-periodic oscillation in the magneto-drag}
\vspace*{1.8cm}
\par\noindent
around zero
with frequency $h\Delta n/2e$, with
$\Delta n$ the density difference between the 2DEGs. Our finding
that the drag is selective to the spin of the electrons is
surprising since established coupling mechanisms via Coulomb
or phonon interactions are a priori not sensitive to spin, as
spin-orbit \hbox{interaction is extremely weak for electrons in GaAs.}
\par
In a drag experiment a current is driven through one of two electrically
isolated layers, the so called drive layer. Interlayer carrier-carrier
scattering through phonons, plasmons or the direct Coulomb interaction
transfers part of the momentum of the carriers in the drive layer to those
in the drag layer, causing a charge accumulation in the drag layer in the
direction of the drift velocity of carriers in the drive layer. The drag
($\rho_T$) is defined as minus the ratio of the electric field originating
from this charge accumulation, to the drive current density. $\rho_T$ of
layers with the same types of carriers is thus expected to be positive,
while that of layers with different types of carriers should be negative.
\par
We have studied transport in several double quantum wells
fabricated from three wafers that differ in the thickness of their
barrier only. The 20 nm wide quantum wells are separated by
Al$_{0.3}$Ga$_{0.7}$As barriers with widths of 30, 60 or 120 nm.
The densities per quantum well are typically 2$\cdot$10$^{11}$ cm$^{-2}$ and
all mobilities exceed
2$\cdot$10$^6$ cm$^2$V$^{-1}$s$^{-1}$. The presented
results are obtained on 30 nm barrier samples, and qualitatively
identical results are obtained on samples
fabricated from the other wafers. Measurements were carried out on
Hall bars with a width of 80 $\mu$m and a length of 880 $\mu$m. Separate
contacts to each quantum well are achieved through the selective depletion
technique~\cite{Eisenstein1} using ex-situ prepared n-doped buried
backgates~\cite{Rubel} and metallic front gates. Measurements were
performed in a $^3$He system with the sample mounted at the end of a cold
finger.
Standard drag-tests (changing ground in the drag layer, interchanging
drag and drive layer, current linearity, and changing the direction of
the applied magnetic field~\cite{Gramila1}) confirmed
that the signal measured is a pure drag-signal.
\par
\begin{figure}
\centering
\begin{turn}{270}
\epsfig{file=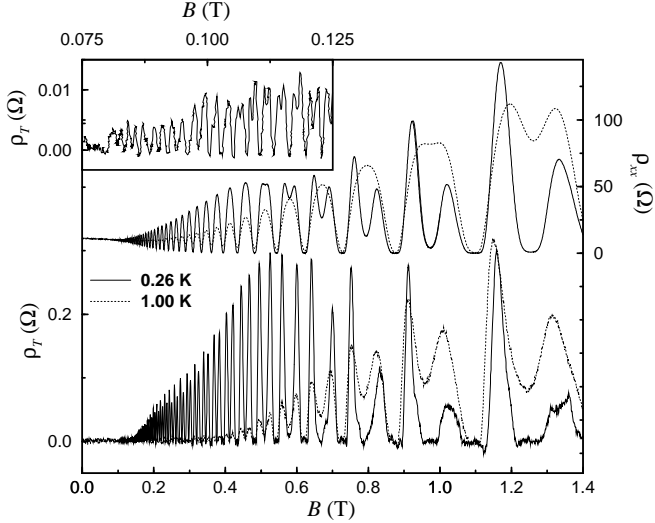, height=0.37\textheight,clip=,bbllx=37,bblly=44,bburx=268,bbury=329}
\end{turn}
\caption{$\rho_{T}$ (bottom) and $\rho_{xx}$ (top) at 0.26~K and 1.0~K and
at matched densities ($n_1$=$n_2$=2.13$\cdot$10$^{11}$ cm$^{-2}$), showing
the absence of a double peak in $\rho_T$ for completely spin-split peaks
in $\rho_{xx}$. Inset is a blow-up of $\rho_T$
at 0.26~K, showing a double peak in fields above 0.11~T.
}
\label{figure1}
\end{figure}

Fig.~\ref{figure1} plots $\rho_T$ and $\rho_{xx}$ measured at temperatures of
\par\noindent
0.26 and 1.0~K. With increasing magnetic field $\rho_{xx}$ shows the usual
Shubnikov-de Haas oscillations which, at
0.26~K, start at a magnetic field of 0.07~T. Spin-splitting becomes visible
at a magnetic field of 0.51~T, and it is completely developed at 1.2~T. By
contrast, at 0.26~K
the oscillations in $\rho_T$ show a double peak in magnetic fields
as low as 0.11~T ($\nu$=77, see inset). The
appearance of a double peak in $\rho_T$ at fields and temperatures
where $\rho_{xx}$ shows no spin-splitting yet has been predicted
theoretically~\cite{Bonsager}. The theory states that $\rho_T$ consists
essentially of the product in the density of states (DOS) at E$_F$ in
each layer, multiplied with the strength of the interlayer
interaction. This strength supposedly strongly decreases at the centre of
a Landau level where, due to the large DOS at E$_F$, screening is
very effective. The decrease would then more than compensate for the increase
in the product of the DOS of the 2DEGs, thus resulting in a double peak
in $\rho_T$. The theory was consistent with experiments
described in a subsequent paper~\cite{Rubel2}. However, the most critical test
for the theory, namely the occurrence of a double peak in $\rho_T$ measured
at a fully spin-split Landau level (that doesn't
show fractional features), could not be performed due to the moderate
mobility of the sample and the accessible temperature range. Our experiment
does allow such a test and fig.~\ref{figure1} shows that $\rho_T$ does
{\em not} show this predicted double peak for spin-split Landau
levels. We further note that at 1~T the longitudinal conductivity in
our sample is 50\% {\em higher} then in the experiment~\cite{Rubel2}
and the theory~\cite{Bonsager} and screening should thus be even
more effective in our samples. The theory is thus not applicable to
explain our experimental results and one is forced to reconsider
the possible role of spin.
\begin{figure}
\centering
\begin{turn}{270}
\epsfig{file=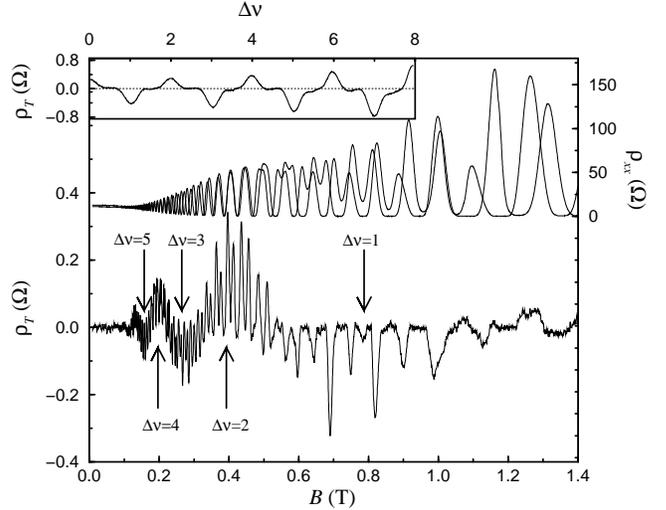, height=0.37\textheight,clip=,bbllx=36,bblly=40,bburx=268,bbury=332}
\end{turn}
\caption{$\rho_T$ (bottom) and $\rho_{xx}$ (top) for both 2DEGs at mismatched
densities ($n_1$=2.27 and $n_2$=2.08$\cdot$10$^{11}$ cm$^{-2}$) as a
function of magnetic field at $T$=0.25 (K). Two sets of oscillations can
be distinguished in $\rho_T$: i) a quick one resulting from
the overlap of the Landau level in the 2DEGs and ii) a slow one which
causes ({\em positive}) maxima in $\rho_T$ whenever the fillingfactor
difference between the 2DEGs is {\em even}, and ({\em negative}) minima
whenever this difference is {\em odd}. The inset shows $\rho_T$ at fixed
magnetic field of 0.641 (T) (maximum in $\rho_T$ in fig.1) versus
fillingfactor difference.
}
\label{figure2}
\end{figure}
\par\noindent
We note furthermore that at 0.11~T and 0.26~K
the bare spin energy (g$\mu_B$B) is only one tenth of the thermal energy
so there is a significant
thermal excitation between the Landau levels with different spin.
This rules out enhanced spin-splitting~\cite{Hill,Nicholas,Fogler}
as the cause for the double peak in $\rho_T$. In the following
we will nonetheless show that it is spin that is
causing the double peak, through a mechanism
where electrons with parallel spin in each layer have a
positive and those with anti-parallel spin have a negative contribution to
$\rho_T$. The minima at large odd fillingfactor then occur, because the
positive and negative contributions cancel.
\par
In order to prove the above scenario
we have measured magneto-drag at mismatched densities. Then
successive Landau levels in the 2DEGs pass through their E$_F$
at different magnetic fields. At certain magnetic fields (depending on
density and density difference of the 2DEGs) the situation
will be such that Landau levels with anti-parallel spins will be at E$_F$
in the 2DEGs, while at somewhat different magnetic fields Landau levels
with parallel spin will be at E$_F$.
Alternatively we have fixed the magnetic field and used one of the gates in
the sample to change the density in one 2DEG, bringing
about the same effect. The first measurement is plotted in the lower
part of fig.~\ref{figure2} together with $\rho_{xx}$ of both
2DEGs (top). As is apparent, for mismatched densities $\rho_T$ is no longer
always positive. Instead $\rho_T$ consists of the sum of two 1/B-periodic
oscillations. A quick one with the frequency $h(n_1+n_2)/2e$, that results
from the overlap of the (in $\rho_T$ for $B>$0.17~T doubly peaked)
Landau levels of the 2DEGs plus a slower one with the frequency
$h(n_1-n_2)/2e$, which causes $\rho_T$ to oscillate around zero. The arrows in
fig.~\ref{figure2} indicate the magnetic fields at which the fillingfactor
difference between the 2DEGs ($\Delta\nu$=$\nu_1$-$\nu_2$) equals an integer. 
$\Delta\nu$ is calculated from the densities of the 2DEGs that are
obtained from the positions of the minima in the Shubnikov-de Haas
oscillations in $\rho_{xx}$. It is clear that when $\Delta\nu$ is {\em odd}
$\rho_T$ is most {\em negative}, while when $\Delta\nu$ is {\em even}
$\rho_T$ is most {\em positive}. The inset of fig.~\ref{figure2} confirms
this even/odd behavior. It plots $\rho_T$ at 0.641~T
($\nu_1=13.5$, maximum $\rho_T$ in fig.~\ref{figure1}) versus $\Delta\nu$
which is changed continuously by decreasing the density of one 2DEG with
a gate. In such a measurement the DOS in the other 2DEG is kept constant, thus
removing the quick oscillation. However, the periodic slow oscillation
with alternating sign still remains and its amplitude increases upon
decreasing the density in the second 2DEG.
\par
The observation of {\em negative} $\rho_T$ at {\em odd} $\Delta\nu$ and
{\em positive} $\rho_T$ at {\em even} $\Delta\nu$ hints the involvement
of spin. If spin-splitting were fully developed, odd $\Delta\nu$ corresponds
to electrons with anti-parallel spin at the E$_F$'s in the 2DEGs. In our
experiment, however, negative $\rho_T$ is observed in
the regime of incomplete spin-splitting. One may then expect a maximum
positive $\rho_T$ at $\Delta\nu$=even and a maximum negative $\rho_T$ at
$\Delta\nu$=even+$\Delta\nu_{spin}$, with $\Delta\nu_{spin}$ the
fillingfactor difference between spin$\uparrow$ and spin$\downarrow$
peaks in $\rho_{xx}$ (which equals 1 only if spin-splitting is
complete). A simulation of $\rho_T$ (see below), assuming positive
coupling between electrons with parallel spins and negative coupling between
electrons with anti-parallel spins, shows however that $\rho_T$ is most
positive for $\Delta\nu=$even and most negative for $\Delta\nu=$odd,
irrespective of the magnitude of the spin-splitting. This magnitude only
influences the amplitude of the oscillations in $\rho_T$, but does
{\em not} alter its phase or periodicity.
\par
\begin{figure}
\centering
\begin{turn}{270}
\epsfig{file=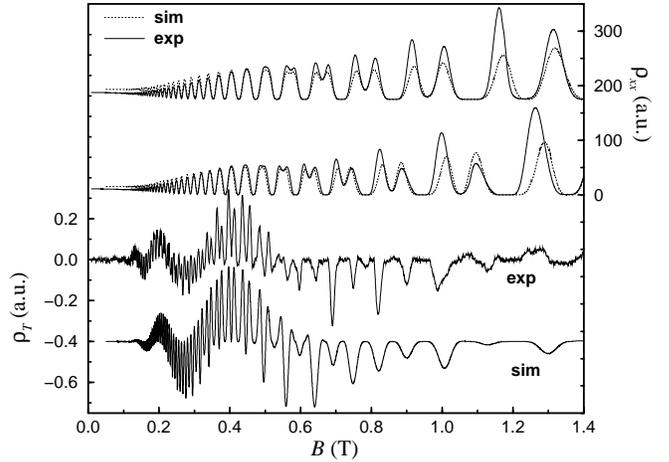, height=0.37\textheight,clip=,bbllx=63,bblly=41,bburx=268,bbury=329}
\end{turn}
\caption{Comparison of simulation and experiment of $\rho_{xx}$ and $\rho_T$
(details can be found in the text). Toptraces show
$\rho_{xx}$, upper curves are offset vertically (solid line: experiment,
dotted line: simulation). Lower traces show the drag, the simulation is
offset vertically.
}
\label{figure4}
\end{figure}

In lack of a theory to compare our results with, we
\hbox{present an empirical model, assuming
$\rho_{xx}\propto$(DOS$^{\uparrow}$+} DOS$^{\downarrow}$)$^2$ and
\hbox{$\rho_T\propto$B$^{\alpha}$(DOS$^{\uparrow}$-DOS$^{\downarrow}$)$_{layer1}\times$(DOS$^{\uparrow}$}-DOS$^{\downarrow}$)$_{layer2}$, with
DOS$^{\uparrow,\downarrow}$ the density of states at E$_F$ for spin$\uparrow$
and spin$\downarrow$, and $B$ the magnetic field. To account for the
unknown change in the coupling between the layers with magnetic field
a factor of $B^{\alpha}$ ($\alpha\approx$-3.5) is used to scale the amplitude of $\rho_T(B)$ to
approximately the experimental value.
The DOS at E$_F$ is given by the sum of a set of Gaussians with
an intrinsic width (due to disorder and temperature) plus a width that
increases with $\sqrt B$. The intrinsic width (1.5~K) is
extracted from the experiment through a Dingle analysis of the
oscillatory part of the low field Shubnikov-de-Haas oscillations. The
coefficient in front of the
$\sqrt B$ (2.7~K for the lower density 2DEG and 2.3~K for the other) is
determined by fitting the simulated $\rho_{xx}$ to the measured one.
In the simulation the densities are kept constant (i.e. E$_F$
oscillates) and \hbox{for the results shown in fig.~\ref{figure4}
we assume an}
\noindent
exchange enhanced spin gap:
$\Delta_{spin}=g\mu_BB+
|(n^{\uparrow}-n^{\downarrow})/(n^{\uparrow}+n^{\downarrow})|\times2E_c$,
with $E_c$ the Coulomb energy $e^2$/$4\pi\epsilon l_B$, $g$ the bare
g-factor in GaAs (-0.44),
$\mu_B$ the Bohr magneton, $\epsilon$ the dielectric
constant, $l_B$ the magnetic length and $n^{\uparrow,\downarrow}$ the
number of particles with spin$\uparrow$ and
spin$\downarrow$. There is some discussion in literature
whether in low fields the relevant length scale for $E_c$ is
$l_B$ or (the much smaller) $k_F^{-1}$ (see~\cite{Nicholas} and
references therein). In our simulation $0.5 l_B$ is appropriate,
i.e. the factor of 2$E_c$ is used as it reproduces the experimental
$\rho_{xx}$ traces. With a fixed enhanced g-factor (or even the
bare g-factor), \hbox{however, qualitatively similar results for $\rho_T$
are obtained.}
\par
Fig.~\ref{figure4} shows the results of the simulation. For both
$\rho_{xx}$ and $\rho_T$, the overall agreement between simulation 
and experiment is satisfactory. For matched densities (not shown)
using the same parameters the agreement is equally good. In fields above
0.8~T the asymmetry
in the height of the experimental spin-split $\rho_{xx}$-peaks is not
reproduced, but this could be due to a different coupling strength of
spin$\uparrow$ and spin$\downarrow$ edge channels to the bulk~\cite{Nachtwei},
which is not included in the simulation. The asymmetry in the $\rho_T$-peaks
at matched densities
(fig.~\ref{figure1}, $B>$0.65~T) may have a similar origin. The simulation
also fails to reproduce some of the finer details in the amplitude of the
quick oscillation in $\rho_T$, but we find that this amplitude is
quite sensitive to overlap between Landau levels in different layers which
in turn depends on details in their width and separation.
\par
The two sets of oscillations in $\rho_T$ are observed in all samples from all
three wafers at mismatched densities. The slow
oscillation can be recognised as such for T$<\sim$1K
although a few negative spikes remain visible till 1.4-1.9~K (depending
on density difference). The inverse period of the slow oscillation is
accurately given by $h/2e \times\Delta n$ in the density range studied
($\Delta n \in [0,1.2] \times 10^{11}$ cm$^{-2}$,
$n_1$=2.0$\times$10$^{11}$ cm$^{-2}$) confirming that the
appearance of negative $\rho_T$ for odd $\Delta\nu$ and positive $\rho_T$ for
even $\Delta\nu$ is not restricted to one particular density difference.
\par
The appearance of {\em negative} $\rho_T$ when Landau levels with anti-parallel
majority-spin are at $E_F$ in the 2DEGs is a puzzling result, as
it implies that electrons in the drag layer gain momentum in the
direction {\em opposite} to that of the net momentum lost by electrons in
the drive layer. In the single particle picture, this can
only occur if the dispersion relation for electrons has a hole-like
character (i.e. $\partial^2 E/ \partial k_{y}^2<0$~\cite{Gramila2}), but
we know of no mechanism through which spins can cause that. The explanation
for negative $\rho_T$ must then be sought for beyond the single particle
picture, possibly in terms of spin-waves or coupled states between the
layers. We note that our empirical formula describing $\rho_T$ consists
of the three possible triplet spin wavefunctions and one could
speculate about an interaction between electrons with opposite momentum in
the different layers. Considering the observation of the effect in the
120 nm barrier samples, the coupling mechanism is most likely not the
direct Coulomb interaction. In any case, our results at least convincingly
demonstrate the importance of the electron spin.
\par
Our empirical model seems to accurately describe $\rho_T$. There is however
a limitation to its applicability: in fields above 1.2 T the negative $\rho_T$
vanishes in the 30 nm barrier samples. For the density mismatch in
fig.~\ref{figure2} this is easily explained, as in fields above $\sim$1.2~T
there is no more chance of finding an overlap between Landau levels with
different spin. However, for larger density differences, such that
there is the necessary overlap of Landau levels with different spin,
we only find positive $\rho_T$ for all temperatures studied
(0.25~K$<$T$<$10~K). We note that at our lowest temperature (0.25~K)
the field of 1.2~T corresponds to a complete spin-splitting in $\rho_{xx}$.
Samples from the other wafers have similar spin-splittings and
the negative $\rho_T$ vanishes at comparable fields. It is further
worth noting that the upper bound for the magnetic field below which
negative $\rho_T$ is observed, does not depend on density or density
difference of the 2DEGs (provided an overlap exists between
Landau levels with different spin for $B>$1.2~T) and thus not
on fillingfactor.
\par
Finally we comment on the interpretation of negative magneto-drag in
ref.~\cite{Gramila2}. Due to the higher lowest temperature (1.15~K),
no spin-splitting in $\rho_{xx}$ and no slow oscillations in $\rho_T$ were
observed. Nevertheless, the remains of half of a slow period which
was filled up with the quick oscillation, were visible. It thus
seemed that negative $\rho_T$ appeared {\em only} when in one 2DEG the
Landau level at E$_F$ was more than half filled, while in the other
the Landau level at E$_F$ was less than half filled. It was argued that
disorder induces a hole-like dispersion in the less-than-half-filled
Landau level, leading to negative $\rho_T$. Our lower temperatures allow
probing the regime where $\rho_{xx}$ shows spin-splitting. The
less-than-half-filled, more-than-half-filled Landau level explanation
should hold for spin-split Landau levels as well, thus doubling the
frequency of the quick oscillation in $\rho_T$. Our experiment shows no
doubling, disproving such a scenario. Moreover, as fig.~\ref{figure2} shows,
negative $\rho_T$ can be observed as well when the (in $\rho_{xx}$ partly or
almost completely spin-split) Landau levels are both less than half filled
(0.62~T, 0.73~T) or both more than half filled (1.0~T). Our data are thus
inconsistent with the interpretation given in ref.~\cite{Gramila2}, while
our empirical model does explain the data of ref.~\cite{Gramila2}.
\par
Summarising, at matched densities the double peak in the
magneto-drag measured at fields and temperatures where the
longitudinal resistance shows no spin-splitting at all,
is the result of the drag being selective to the spin of the electrons,
such that electrons with parallel spin in each layer have a positive
contribution to the drag, while those with anti-parallel spin have a negative
contribution. This selectivity to spin further causes the occurrence of a
{\em negative} drag whenever Landau levels with anti-parallel spin are at
$E_F$ in the 2DEGs, resulting in a novel 1/B-periodic oscillation
in the low field low temperature drag for mismatched electron densities
with the inverse period given by $h\Delta n/2e$. Our empirical model
assuming $\rho_T\propto$ (DOS$^{\uparrow}$-DOS$^{\downarrow}$)$_{layer1}\times$(DOS$^{\uparrow}$-DOS$^{\downarrow}$)$_{layer2}$ quite accurately describes
the results at matched as well as mismatched densities. The origin of the
negative coupling between electrons with anti-parallel spin as well as its
disappearance when spin splitting in $\rho_{xx}$ is complete remains to
be explained.
\par
We acknowledge financial support from BMBF and European Community TMR network
no. ERBFMRX-CT98-0180. We are grateful to L.I.~Glazman for helpful
discussion and to J. Schmid for experimental support.

\end{document}